\pdfoutput=1
\documentclass[sigconf]{acmart}

\AtBeginDocument{%
  }

\setcopyright{none}
\setcctype{by}
\copyrightyear{2026}
\acmYear{2026}
\acmConference[WWW Companion '26] {Companion Proceedings of the ACM Web Conference 2026}{April 13--17, 2026}{Dubai, United Arab Emirates.}
\acmBooktitle{Companion Proceedings of the ACM Web Conference 2026 (WWW Companion '26), April 13--17, 2026, Dubai, United Arab Emirates}

\usepackage{graphicx}
\usepackage{subcaption}
\usepackage{multicol}

\begin{document}
\title{Learning to Alleviate Familiarity Bias in Video Recommendation}





\author{Zheng Ren}
\affiliation{%
  \institution{Google LLC}
  \country{USA}}
\email{murphyren@google.com}

\author{Yi Wu}
\affiliation{%
  \institution{Google LLC}
  \country{USA}}
\email{wuyish@google.com}

\author{Jianan Lu}
\affiliation{%
  \institution{Google LLC}
  \country{USA}}
\email{jiananlu@google.com}

\author{Acar Ary}
\affiliation{%
  \institution{Google LLC}
  \country{USA}}
\email{aryacar@google.com}

\author{Yiqu Liu}
\affiliation{%
  \institution{Google LLC}
  \country{USA}}
\email{yiquliu@google.com}

\author{Li Wei}
\affiliation{%
  \institution{Google LLC}
  \country{USA}}
\email{liwei@google.com}

\author{Lukasz Heldt}
\affiliation{%
  \institution{Google LLC}
  \country{USA}}
\email{heldt@google.com}


\renewcommand{\shortauthors}{Zheng Ren et al.}


\begin{abstract}
Modern video recommendation systems aim to optimize user engagement and platform objectives, yet often face structural exposure imbalances caused by behavioral biases. In this work, we focus on the post-ranking stage and present LAFB (Learning to Alleviate Familiarity Bias), a lightweight and model-agnostic framework designed to mitigate familiarity bias in recommendation outputs. LAFB models user–content familiarity using discrete and continuous interaction features, and estimates personalized debiasing factors to adjust user rating prediction scores, thereby reducing the dominance of familiar content in the final ranking. We conduct large-scale offline evaluations and online A/B testing in a real-world recommendation system, under a unified serving stack that also compares LAFB with deployable popularity-oriented remedies. Results show that LAFB increases novel watch-time share and improves exposure for emerging creators and overall content diversity, while maintaining stable overall watch time and short-term satisfaction. LAFB has already been launched in the post-ranking stage of YouTube’s recommendation system, demonstrating its effectiveness in real-world applications.
\end{abstract}




\maketitle

\section{Introduction}

Modern video platforms serve billions of users through personalized recommendation systems~\cite{10.1145/2959100.2959190}, shaping both user engagement and the visibility of emerging creators~\cite{fairnesses}. While these systems optimize metrics such as click-through rate and watch time, they often suffer from structural biases. One prominent example is familiarity bias: users tend to rate familiar items higher, which in turn causes the system to predict higher ratings for familiar items~\cite{Popularity,li2023fairnessrecommendationfoundationsmethods}, resulting in repetitive recommendations and reduced diversity~\cite{cad}.

In operational systems, mitigation has typically acted at one of three locations in the pipeline: at serving time through exposure rebalancing, inside learned representations via interpretable post hoc attenuation, and through lightweight adjustments grounded in global popularity statistics. User- and item-centric re-ranking recalibrates exposure without modifying the underlying models, and user studies report that overall list satisfaction does not decrease while perceived discovery can increase under item-centric configurations~\cite{recsys2024_user_item}. Interpretable post hoc approaches attenuate popularity amplification by exposing steerable latent factors with sparse autoencoders, enabling model-agnostic mitigation with limited accuracy loss~\cite{recsys2025_interpretable_popbias}. Complementing these, probabilistic analyses motivate a lightweight baseline that adjusts relevance according to global popularity, offering a tunable trade-off between accuracy and long-tail exposure using only aggregate statistics~\cite{crowd_prob_analysis_2024}. These advances largely regulate exposure with respect to global popularity; by contrast, familiarity bias is user- and context-specific. Production-ready mitigation, therefore, benefits from a personalized, low-intrusion mechanism that can accommodate multiple continuous familiarity signals and remain stable under behavioral shifts.

We propose Learning to Alleviate Familiarity Bias (LAFB), a lightweight framework applied at the post-ranking stage. LAFB directly targets familiarity bias by learning personalized debiasing factors from user feedback, derived from both discrete and continuous familiarity features, and then adjusting a model’s rating-based signal before it enters the final ranking logic. In contrast to generic re-ranking or static popularity penalties~\cite{10.1145/3097983.3098173,inproceedings,bm2022}, LAFB leaves the rest of the system unchanged while adapting to evolving user behavior.

LAFB has been successfully deployed in the post-ranking stage of YouTube’s recommendation system, demonstrating its effectiveness in large-scale real-world applications.

Our contributions are summarized as follows:
\begin{itemize}
\item We propose LAFB, a lightweight post-ranking framework that calibrates URPS using debiasing factors learned from both discrete bucketed and continuous familiarity features.
\item We establish a unified evaluation on a single serving stack, introduce the Novel WT Share metric and a calibrated score-distribution view, and compare against deployable popularity-oriented remedies (user-/item-centric re-ranking, interpretable SAE, and log-pop penalization).
\item We validate at scale with offline studies and online A/B on production traffic, showing reduced familiar-share, higher novel consumption, and improved emerging-creator exposure while maintaining overall watch time; LAFB is deployed in YouTube post-ranking.
\end{itemize}



\section{Methodology}
We start with a User Rating Prediction Score (URPS) \( s_{u,v} \), which represents the predicted preference of user \( u \) for item \( v \). URPS can come from any model that estimates user--item satisfaction rating. It is commonly observed that users tend to give higher ratings to familiar items, causing a familiarity bias in model predictions trained on such data.

Given the URPS \( s_{u,v} \) for user \( u \) and item \( v \), and a set of familiarity features \( b_{u,v} = (b_{u,v}^1, b_{u,v}^2, \ldots, b_{u,v}^n) \), such as historical interaction counts (e.g., watch count, click count) and recency of last interaction, we aim to remove familiarity bias by decorrelating the prediction score \( s_{u,v} \) from the familiarity features \( b_{u,v} \).

We define the debiased score as 
\begin{equation}
  s^{\mathrm{debias}}_{u,v} = \frac{s_{u,v}}{E[s_{u,v} \mid b_{u,v}]},
  \label{eq:debias}
\end{equation}

By conditional expectation,

\begin{equation*}
  E[s^{\mathrm{debias}}_{u,v} \mid b] = 1.
\end{equation*}
Given that the expected value of $s^{\mathrm{debias}}_{u,v}$ is a constant for every realization of $b$, this also implies $\operatorname{Corr}(b_{u,v}^i, s^{\mathrm{debias}}_{u,v}) = 0$ for every $i\in [n]$.

\subsection{Estimating Debiasing Factor}
Given URPS score \( s_{u,v} \), we would like to compute the debiasing factor $E[s_{u,v} \mid b_{u,v}]$. This can be achieved by either using empirical mean or building a functional approximator.
\paragraph{Discrete Feature Bucketization.}

For low-dimensional categorical familiarity features, we use the empirical mean of \( s_{u,v} \) as the debiasing factor. Specifically, for each feature bucket \( b = (b_1, \dots, b_n) \), we compute:

\begin{equation*}
  \text{Adj}_b = 
  \frac{\sum_{\{(u,v):\,(b_{u,v,1}, \dots, b_{u,v,n}) = b\}} s_{u,v}}
       {\lvert \{(u,v):\,(b_{u,v,1}, \dots, b_{u,v,n}) = b\} \rvert},
  \label{eq:multi_discrete_debias}
\end{equation*}

\begin{figure*}[htbp]
  \centering
  \begin{subfigure}[t]{0.2\textwidth}
    \centering
    \includegraphics[width=\linewidth]{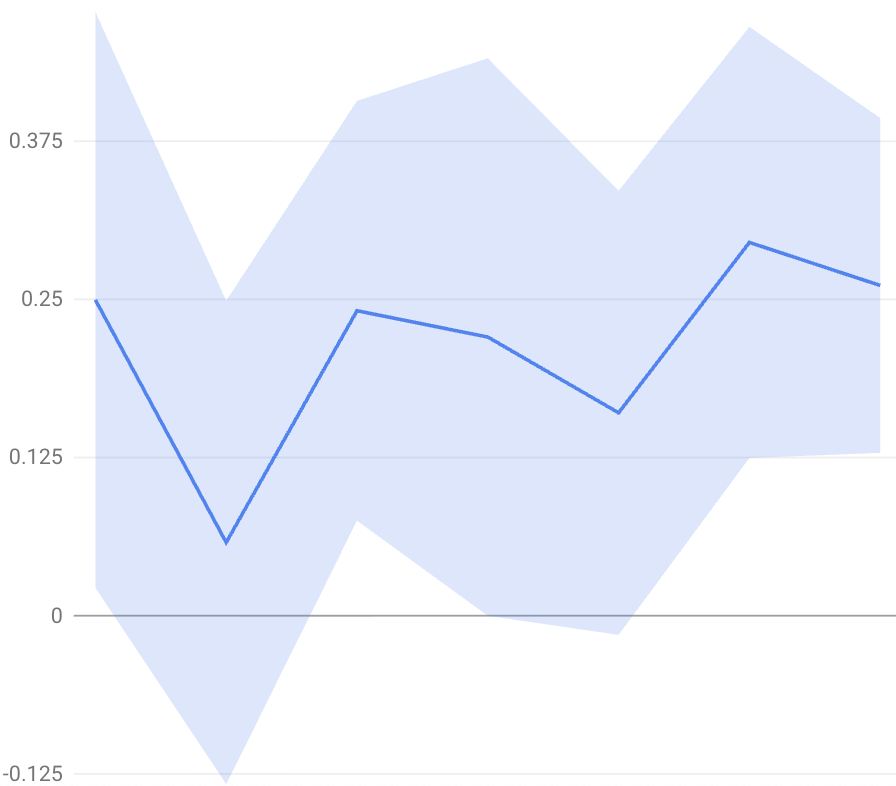}
    \caption{Creator Exposure}
  \end{subfigure}
  \hfill
  \begin{subfigure}[t]{0.2\textwidth}
    \centering
    \includegraphics[width=\linewidth]{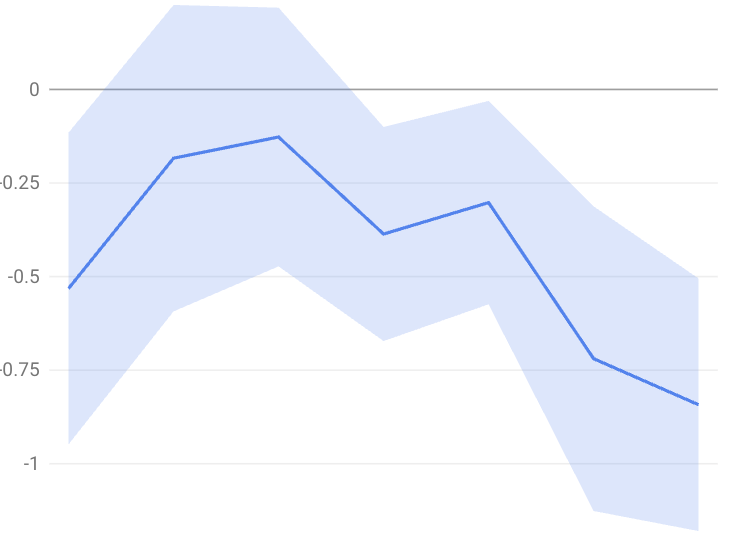}
    \caption{Familiarity Suppression}
  \end{subfigure}
  \hfill
  \begin{subfigure}[t]{0.2\textwidth}
    \centering
    \includegraphics[width=\linewidth]{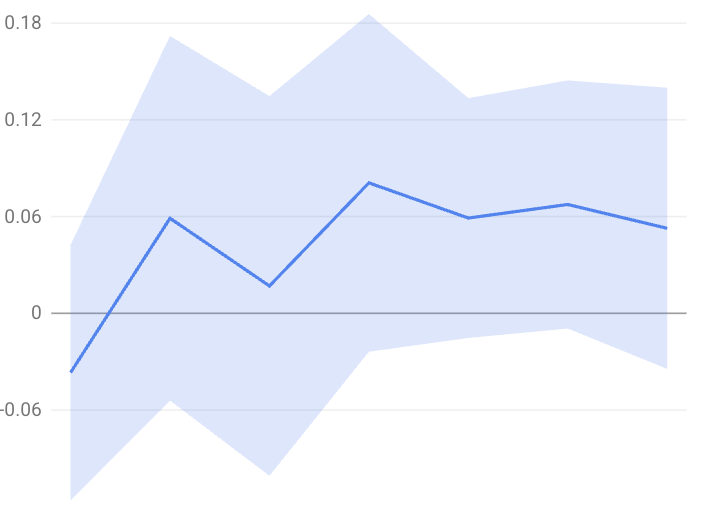}
    \caption{Overall WT}
  \end{subfigure}
  \hfill
  \begin{subfigure}[t]{0.2\textwidth}
    \centering
    \includegraphics[width=\linewidth]{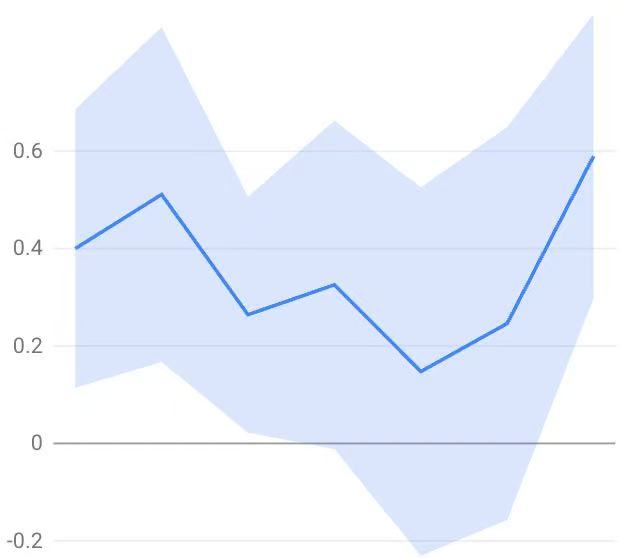}
    \caption{Novel WT}
  \end{subfigure}
  \caption{Performance under multi-discrete familiarity modeling.}
  \label{fig:discrete-multi}
\end{figure*}

\begin{figure*}[htbp]
  \centering
  \begin{subfigure}[t]{0.2\textwidth}
    \centering
    \includegraphics[width=\linewidth]{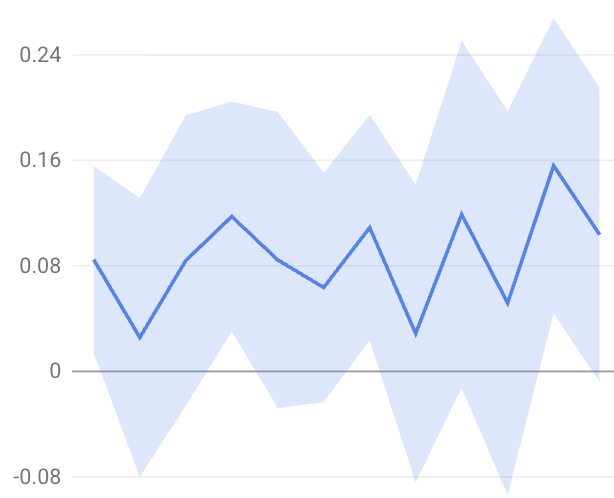}
    \caption{Creator Exposure}
  \end{subfigure}
  \hfill
  \begin{subfigure}[t]{0.2\textwidth}
    \centering
    \includegraphics[width=\linewidth]{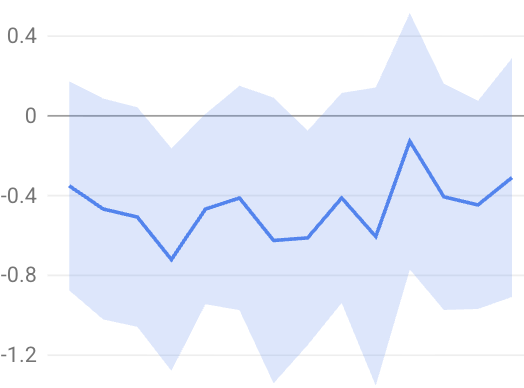}
    \caption{Familiarity Suppression}
  \end{subfigure}
  \hfill
  \begin{subfigure}[t]{0.2\textwidth}
    \centering
    \includegraphics[width=\linewidth]{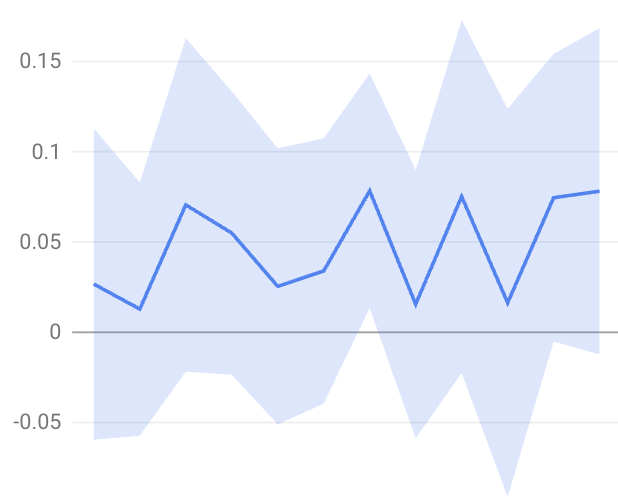}
    \caption{Overall WT}
  \end{subfigure}
    \hfill
  \begin{subfigure}[t]{0.2\textwidth}
    \centering
    \includegraphics[width=\linewidth]{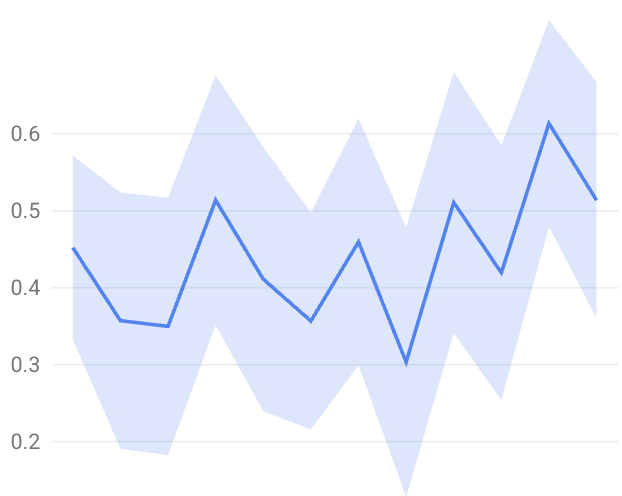}
    \caption{Novel WT}
  \end{subfigure}
  \caption{Performance under multi-continuous familiarity modeling.}
  \label{fig:continuous-multi}
\end{figure*}

\paragraph{Continuous Feature Modeling.}

For high-dimensional continuous familiarity features, we use a small neural network \( f(b; \theta) \), where \( b = (b_1, b_2, \dots, b_n) \) denotes the continuous familiarity feature vector. The network is trained to predict the corresponding user rating prediction score \( s_{u,v} \) using mean squared error (MSE) loss; i.e., we minimize the following objective function:

\begin{equation*}
\min_{\theta} \; \mathbb{E}_{(u,v)} \left[ \left( f(b_{u,v}; \theta) - s_{u,v} \right)^2 \right],
\end{equation*}

Once trained, the debiasing factor for any feature vector \( b \) is computed as $\mathrm{Adj}_b = f(b; \theta)$.

\subsection{Final Score Adjustment}
Given a user \( u \) and item \( v \) with User Rating Prediction Score (URPS) \( s_{u,v} \), we normalize the original score by dividing it by the estimated debiasing factor. The debiased prediction score is computed as:

\begin{equation*}
s^{\mathrm{debias}}_{u,v} = \frac{s_{u,v}}{\mathrm{Adj}_{b_{u,v}}}.
\end{equation*}

In many recommendation systems, a ranking function \( R \) computes a real-valued score used to order items. It typically takes a single URPS \( s_{u,v} \), or multiple URPS scores, along with other quality signals $X_{u,v}$ as input. We propose to replace each URPS with its debiased version \( s^{\mathrm{debias}}_{u,v} \), leading to:

\begin{equation}
\mathrm{Score}_{u,v} = R(s^{\mathrm{debias}}_{u,v}, X_{u,v}),
\end{equation}

This replacement mitigates familiarity bias while preserving the core structure of the existing ranking function.

\begin{figure}[htbp]
  \centering
  \subfloat[Before debiasing]{\includegraphics[width=0.45\linewidth]{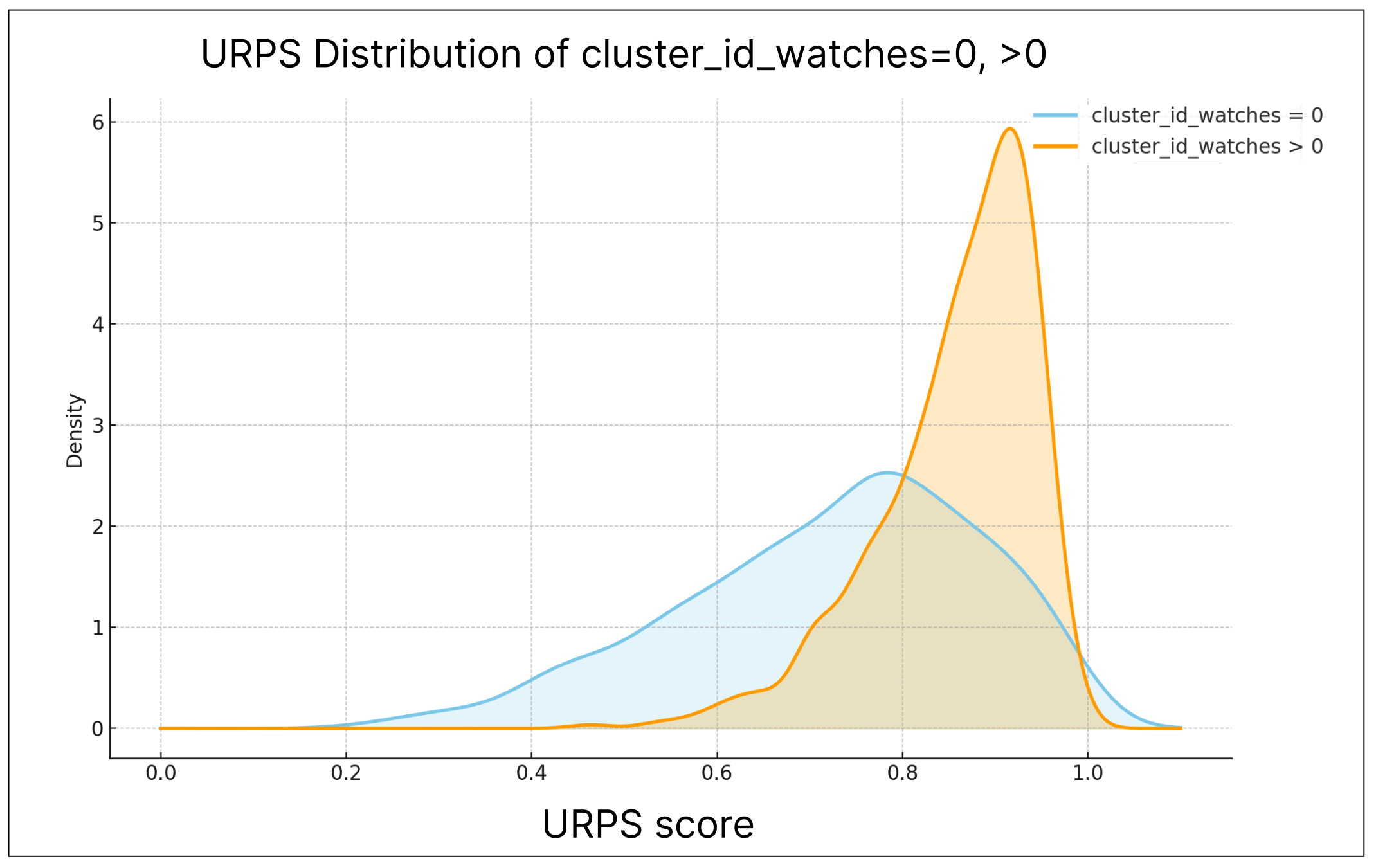}}
  \hfill
  \subfloat[After debiasing]{\includegraphics[width=0.45\linewidth]{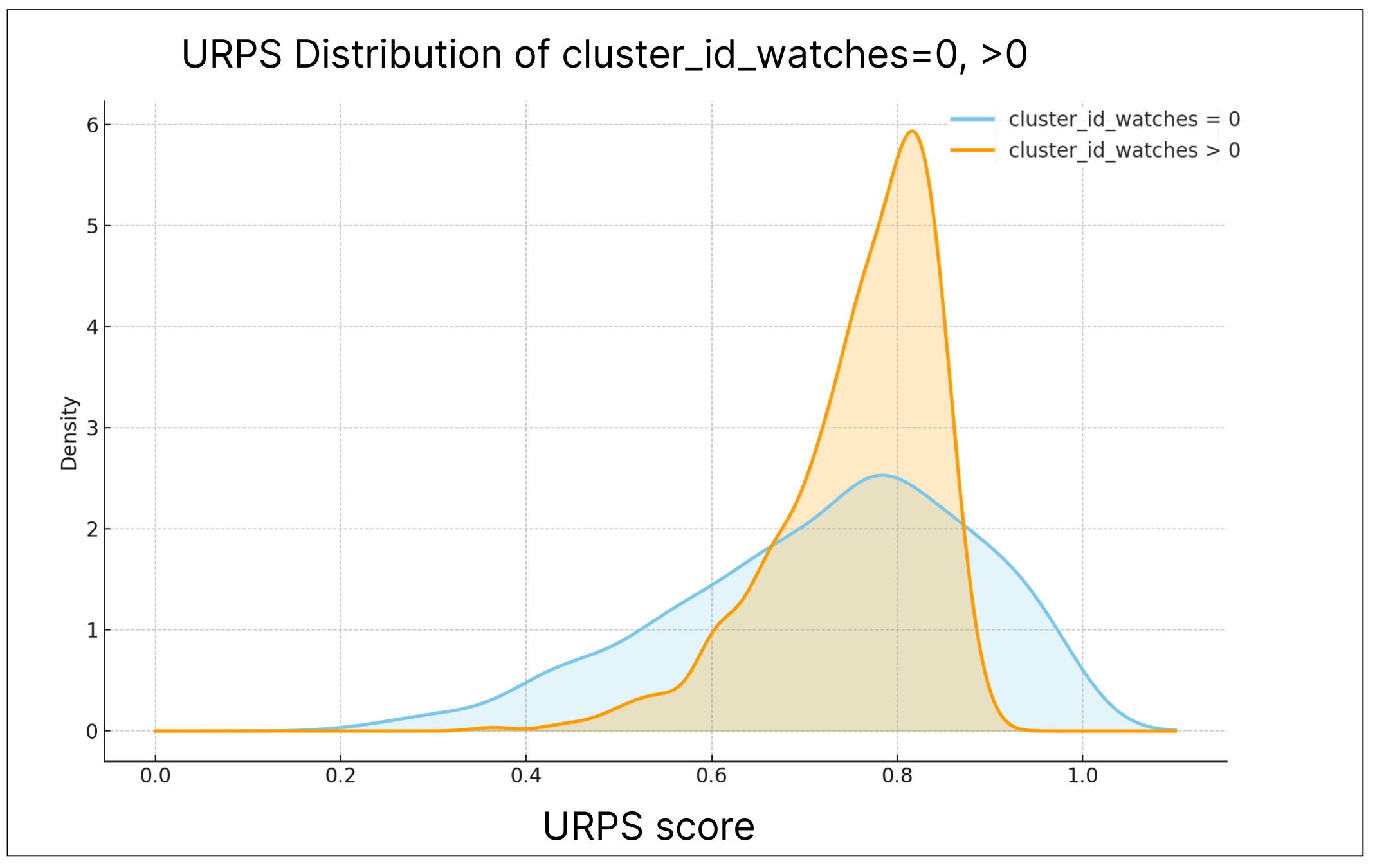}}
  \caption{URPS score distributions across fine-grained familiarity levels (low, medium, and high), shown before (a) and after (b) debiasing.}
  \label{fig:score-distribution}
\end{figure}

\subsection{Comparison with Heuristic Boost}

Many existing recommendation systems mitigate familiarity bias using manual boosting: fixed score adjustments applied to unfamiliar content based on predefined rules. While simple, this approach has two key limitations:

\begin{itemize}
    \item \textbf{Simple-Feature Limitation.} Manual boosting usually relies on one or two discrete features, ignoring richer continuous familiarity signals and their interactions.
    \item \textbf{No Adaptivity.} Boost coefficients remain static, unable to adjust as user behavior or platform dynamics change.
\end{itemize}

LAFB addresses both gaps by learning debiasing factors automatically from user feedback data. It supports multiple discrete and continuous features and continuously adapts correction strength as user behavior evolves.
\section{Evaluation}
LAFB has been launched in YouTube production systems for various URPS scores predicting user satisfaction. Here we report results from A/B tests conducted during the system launch.

\subsection{Debiasing with Discrete and Continuous Familiarity Features}



We evaluate LAFB under both discrete and continuous familiarity modeling, using the same serving stack and guardrails across variants. In the discrete setting, interaction features related to familiarity (e.g., watch frequency, creator/genre affinity, and recency) are bucketed and each bucket receives a learned debiasing factor. Figure~\ref{fig:discrete-multi} summarizes the effects. Creator exposure rises in a steady but conservative manner across buckets, while familiar-content watch time is suppressed below the control level for most buckets. Overall watch time remains neutral to slightly positive, indicating that the exposure shift does not compromise short-term engagement when clipping and smoothing are applied. The newly introduced Novel WT metric shows a concurrent lift: the share of watch time allocated to content that is novel for the user increases in step with the reduction in familiar share. The confidence band is wider in sparsely populated buckets, which explains mild oscillations; nonetheless the overall pattern is consistent—more exploration with stable engagement.

In the continuous setting, raw interaction statistics form a multivariate familiarity vector, and the debiasing factor is learned as a smooth function over this space. Figure~\ref{fig:continuous-multi} shows stronger effects along the same directions. Creator exposure gains are larger and more uniform, familiar-content share exhibits a clearer and deeper suppression, and overall watch time shifts slightly upward. Novel WT increases more decisively than in the discrete case, with a narrower confidence band due to cross-bucket sharing inherent to the continuous estimator. Taken together, the continuous variant converts the attenuation of familiar consumption into a broader acceptance of previously unseen items, while keeping engagement stable.

These two views lead to a coherent interpretation. Both variants mitigate familiarity amplification without harming overall watch time, and both translate familiar suppression into higher Novel WT rather than merely reallocating traffic among already-familiar items.

\subsection{Score Distribution Analysis}


We analyze URPS score distributions before and after debiasing across discrete familiarity features, with watch count within the channel bucketed (Figure~\ref{fig:score-distribution}). Before adjustment, scores are concentrated on familiarity content, reinforcing bias. After applying LAFB, the distribution flattens, reducing variance and narrowing the gap between high- and low-familiarity items. This indicates improved fairness while maintaining score smoothness and ranking stability.

\begin{figure*}[htpb]
  \centering
  \begin{subfigure}[t]{0.3\textwidth}
  \centering
  \includegraphics[width=\linewidth]{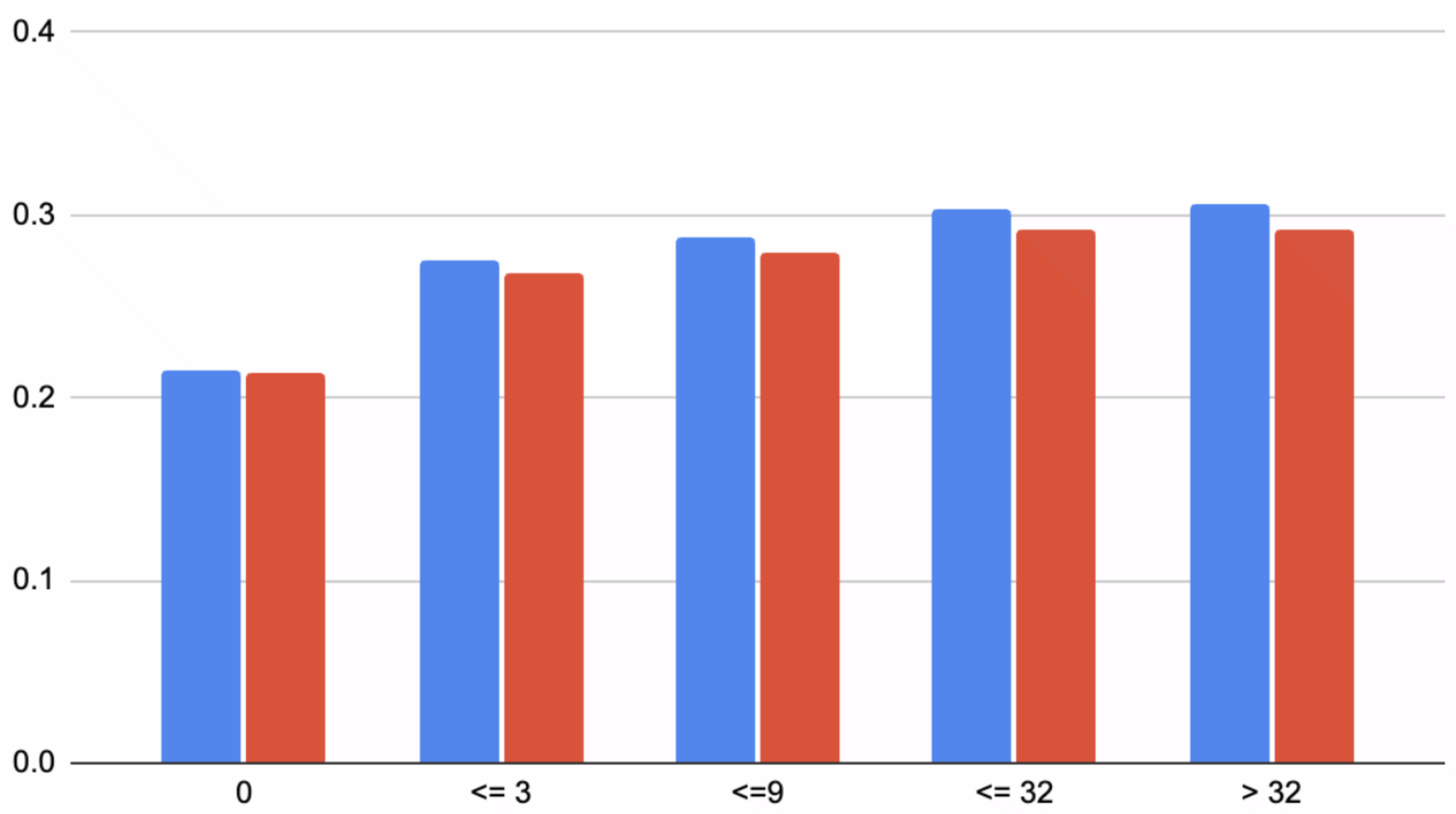}
  \caption{Final Ranking Score Label Before/After the Debias}
  \vspace{0.5\baselineskip} 
  \begin{tabular}{r l}
    \textcolor{blue}{$\blacksquare$} & $s^{\mathrm{debias}}_{u,v}$ (Training Label) \\
    \textcolor{red}{$\blacksquare$} & $s_{u,v}$ (Training Label After the Debias)
  \end{tabular}
\end{subfigure}
  \hfill
  \begin{subfigure}[t]{0.3\textwidth}
  \centering
  \includegraphics[width=\linewidth]{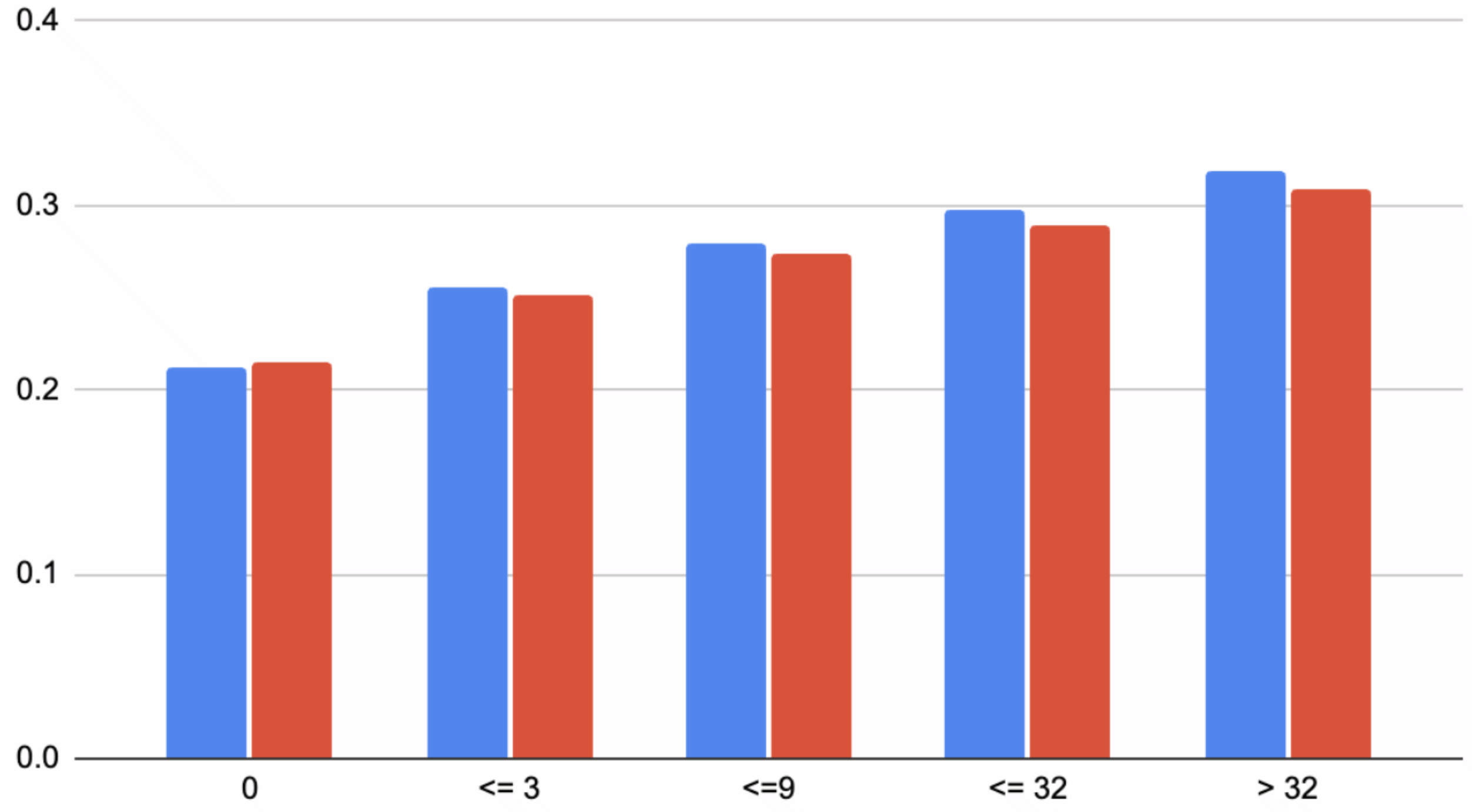}
  \caption{Final Ranking Predict Score Before/After the Debias}
  \vspace{0.5\baselineskip} 
  \begin{tabular}{r l}
    \textcolor{blue}{$\blacksquare$} & $s^{\mathrm{debias}}_{u,v}$ (Predict Score) \\
    \textcolor{red}{$\blacksquare$} & $s_{u,v}$ (Predict Score After the Debias)
  \end{tabular}
\end{subfigure}
  \hfill
   \begin{subfigure}[t]{0.3\textwidth}
  \centering
  \includegraphics[width=\linewidth]{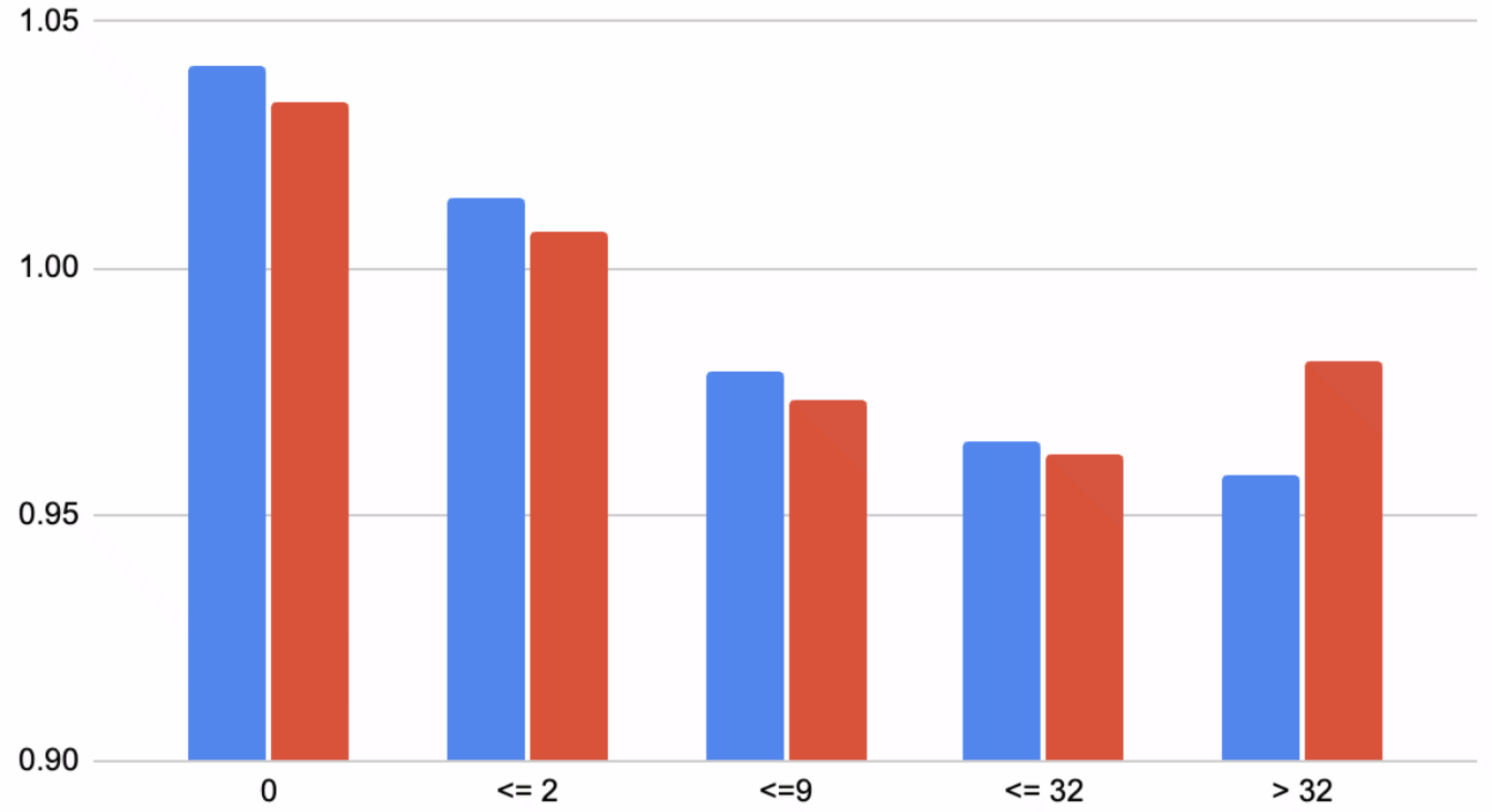}
  \caption{Debias Model Calibration}
  \vspace{0.5\baselineskip} 
  \begin{tabular}{r l}
    \textcolor{blue}{$\blacksquare$} & Before Debias \\
    \textcolor{red}{$\blacksquare$} & After the Debias
  \end{tabular}
\end{subfigure}
  \caption{Comparison of Model Predictions.}
  \label{fig:debiasing_comparison}
\end{figure*}

\subsection{Comparison with other familiarity-oriented remedies}

We compare LAFB with deployable other familiarity-oriented remedies under the same serving stack and traffic split. All interventions operate post-ranking to keep candidate generation and latency unchanged. Metrics are reported as changes versus the production control: Emerging Creator Exposure (relative change for creators flagged as emerging by recent global percentiles), Novel WT Share (pp change for per-user novel content within a 14-day window), Familiar WT Share (pp change for familiar content; lower is better), and Overall WT (percent change in user-day aggregated watch time). Confidence intervals are 95\% with variance reduction using historical covariates.

\begin{table*}[t]
\centering
\small 
\setlength{\tabcolsep}{6pt}
\renewcommand{\arraystretch}{1.15}
\caption{Comparison across watch-time and diversity metrics (online A/B). Values are changes vs. Control; arrows indicate preferred direction.}
\label{tab:main_ab}
\resizebox{\textwidth}{!}{%
\begin{tabular}{lcccc}
\toprule
Method
& Emerging Creator Exposure $\Delta$\% $\uparrow$
& Novel WT Share $\Delta$pp $\uparrow$
& Familiar WT Share $\Delta$pp $\downarrow$
& Overall WT $\Delta$\% $\uparrow$ \\
\midrule
User-centric re-ranking
& +0.01 [ $-0.01$, $+0.03$ ]
& +0.10 [ $-0.05$, $+0.25$ ]
& $-0.08$ [ $-0.20$, $+0.02$ ]
& +0.00 [ $-0.02$, $+0.02$ ] \\
Item-centric re-ranking
& +0.02 [ $-0.01$, $+0.04$ ]
& +0.18 [ $-0.03$, $+0.32$ ]
& $-0.22$ [ $-0.30$, $-0.10$ ]
& +0.02 [ $-0.01$, $+0.04$ ] \\
Interpretable post hoc (SAE)
& $-0.02$ [ $-0.03$, $+0.01$ ]
& +0.86 [ $+0.72$, $+1.00$ ]
& $-0.35$ [ $-0.45$, $-0.20$ ]
& +0.03 [ $-0.01$, $+0.05$ ] \\
Log-pop penalization
& +0.01 [ $-0.01$, $+0.02$ ]
& +0.31 [ $-0.04$, $+0.50$ ]
& $-0.41$ [ $-0.55$, $-0.25$ ]
& +0.02 [ $-0.02$, $+0.03$ ] \\
LAFB (ours)
& \textbf{+0.03 [ $-0.01$, $+0.02$ ]}
& \textbf{+0.83 [ $-0.08$, $+0.10$ ]}
& \textbf{-0.63 [ $-0.04$, $+0.06$ ]}
& \textbf{+0.05 [ $-0.02$, $+0.03$ ]} \\
\bottomrule
\end{tabular}%
}
\end{table*}

Overall, WT is preserved across all arms; LAFB shows a small positive change, confirming that per-user familiarity mitigation does not compromise short-term engagement when clipped and smoothed. Novel WT Share increases most for SAE and LAFB: SAE broadens acceptance by attenuating popularity-related factors in the representation, and LAFB releases exploration capacity by reducing the coupling between score and user-specific familiarity. Familiar WT Share declines for all remedies that promote novelty or tail exposure, with the largest and most stable reduction from LAFB; SAE is second, while item-centric and log-pop offer moderate uniform reductions. Emerging Creator Exposure rises directly under item-centric and log-pop due to redistribution, and rises more modestly yet consistently under LAFB, indicating benefits to emerging creators without overcompensation.

LAFB yields larger gains in Novel WT Share for new or light users and stronger familiar-share suppression for experienced or heavy users; by creator size, item-centric and log-pop mainly transfer head to tail, whereas LAFB improves mid–tail coverage more evenly. Sensitivity sweeps show the usual frontiers: excessive global strength harms Overall WT for item-centric and log-pop or erodes relevance for SAE; LAFB maintains a broad operating region where familiar-share falls and novel-share rises while Overall WT remains neutral to slightly positive. Mechanistic checks confirm action at the intended layer: the correlation between scoring signal and measured familiarity features drops under LAFB, while SAE reduces the energy of popularity-related latent factors; correlation shifts are minor for exposure-only remedies.

\subsection{Performance Evaluation of the Debiasing Framework}


Figure~\ref{fig:debiasing_comparison} provides a critical comparison of model performance across the debiasing process, where the x-axis consistently represents continuous familiarity features (the count of previous watches from a channel), segmented into five equal-sized buckets, each holding $20\%$ of the data distribution. Subfigure (a) illustrates the effect on the training data's final predicted score label and (b) shows the impact on the model's predicted score. In both cases, a key observation confirms the debiasing effectiveness: for the higher-familiarity buckets (larger x-values, meaning content watched many times before), both the debiased label and the debiased predicted score are lower than the original values. This indicates a successful reduction of the inherent popularity bias. Finally, subfigure (c) depicts the Debias Model Calibration, defined as the ratio of the debias factor predicted score to the debias factor label. After incorporating the continuous familiarity features, the calibration improves for three out of five buckets, moving them closer to the ideal value of 1, thereby confirming that the framework enhances content diversity and reduces bias without sacrificing the predictive quality or reliability of the final ranking scores.

\section{Conclusion}

We present LAFB, a lightweight post-ranking debiasing framework for mitigating familiarity bias in recommendation systems. It adjusts quality scores using sample-specific factors from discrete and continuous familiarity features, and integrates seamlessly without modifying the base model. LAFB increases exposure to less familiar content while preserving personalization. Both offline simulations and large-scale online A/B tests confirm its effectiveness in reducing repetition and improving diversity without sacrificing user engagement.



\bibliographystyle{ACM-Reference-Format}
\bibliography{sample-base}


\end{document}